\documentclass[10pt,twocolumn,floatfix,superscriptaddress,nature,longbibliography,notitlepage]{revtex4-1}
\setcitestyle{super}
\usepackage{amsmath,amssymb,amsthm,mathrsfs,amsfonts,dsfont,amstext} 
\usepackage{textcomp,pbox}
\usepackage[export]{adjustbox}
\usepackage{bm}
\usepackage{dcolumn,booktabs,url}
\usepackage[scaled]{helvet}
\usepackage{sansmath,gensymb}
\usepackage{tikz,graphicx,transparent,color}
\usepackage{multirow}
\usepackage[separate-uncertainty = true]{siunitx}
\usepackage{comment}
\usepackage{physics}
\usepackage{pdfpages}
\usepackage{float}
\usepackage{lineno}
\makeatletter
\AtBeginDocument{\let\LS@rot\@undefined}
\makeatother

\usepackage[colorlinks=true]{hyperref}
\usepackage{graphicx}
\hypersetup{
     colorlinks   = true,
     citecolor    = red,
     linkcolor    = blue,
     urlcolor     = red     
}

\usepackage{setspace}

\newcommand{\yso}{Y$_2$SiO$_5$}
\newcommand{\eu}[0]{Eu$^{3+}$}

\newcommand{\euiso}[0]{$^{151}$Eu$^{3+}$:Y$_2$SiO$_5$}

\begin{document}

\newcommand{\TitleName}{Storage of photonic time-bin qubits for up to 20 ms in a rare-earth doped crystal}
\title{\TitleName}

\newcommand{\AffGeneve}{Département de Physique Appliquée, Université de Genève, Rue de l’Ecole-de-Médecine 20, 1205, Genève, Switzerland}
\newcommand{\AffNice}{Université côte d’Azur, CNRS, Institut de Physique de Nice, Parc Valrose, Nice, Cedex 2, France}

\author{Antonio Ortu} 
\affiliation{\AffGeneve{}}
\author{Adrian Holz\"{a}pfel} 
\affiliation{\AffGeneve{}}
\author{Jean Etesse} 
\affiliation{\AffNice{}}
\author{Mikael Afzelius}\email[Corresponding author, ]{mikael.afzelius@unige.ch}
\affiliation{\AffGeneve{}}

\date{\today}

\begin{abstract}
Long-duration quantum memories for photonic qubits are essential components for achieving long-distance quantum networks and repeaters. The mapping of optical states onto coherent spin-waves in rare earth ensembles is a particularly promising approach to quantum storage. However, it remains challenging to achieve long-duration storage at the quantum level due to read-out noise caused by the required spin-wave manipulation. In this work, we apply dynamical decoupling techniques and a small magnetic field to achieve the storage of six temporal modes for 20, 50 and 100 ms in a \euiso{} crystal, based on an atomic frequency comb memory, where each temporal mode contains around one photon on average. The quantum coherence of the memory is verified by storing two time-bin qubits for 20 ms, with an average memory output fidelity of $F=\SI{85\pm2}{\percent}$ for an average number of photons per qubit of $\mu_\text{in}$ = 0.92$\pm$0.04. The qubit analysis is done at the read-out of the memory, using a type of composite adiabatic read-out pulse we developed.
\end{abstract}

\maketitle
\thispagestyle{plain}



\section*{Introduction}
\begin{figure*}[htp]
	\includegraphics[width=\linewidth]{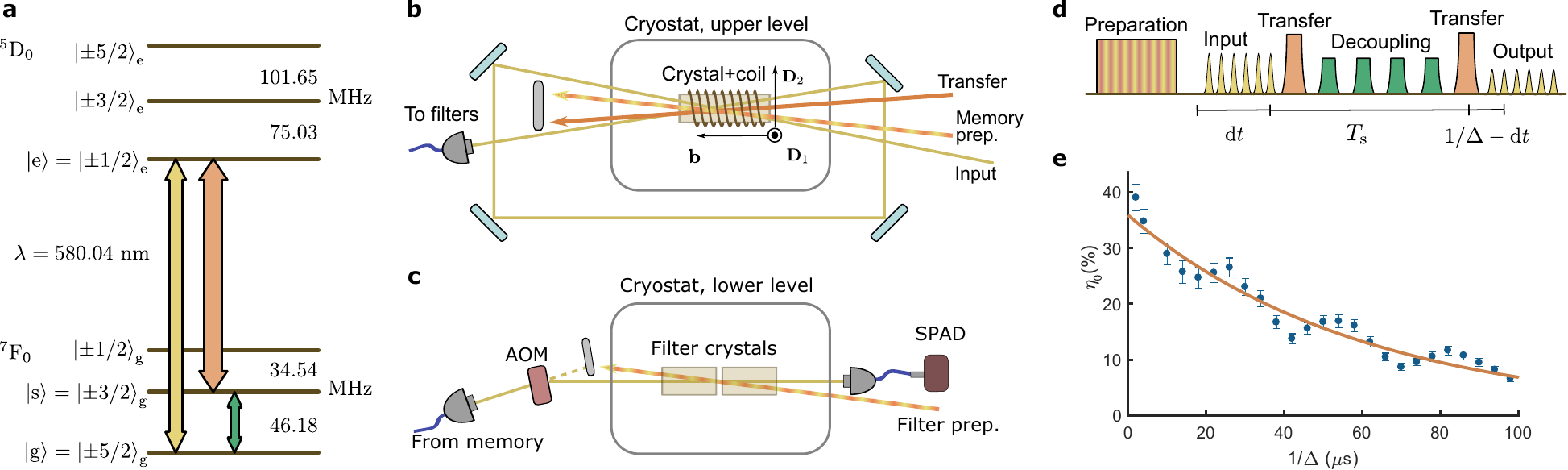}
	\caption{\textbf{System and setup} (a) Atomic energy structure of \euiso{} and transitions used in the memory protocol. (b) and (c), sketches of the experimental setup around the memory and filter crystals, respectively. The crystals are glued on a custom mount with two levels at different heights, in the same cryostat. The memory crystal is at the center of a small coil used to generate the RF signal. A larger coil (not shown) on top of the cryostat generates a static magnetic field along the $\text{\textbf{D}}_1$ axis of the \yso{} crystal. Optical beams are depicted with exaggerated angles for clarity. AOM: acousto-optic modulator; SPAD: single photon detector (d) Sketch of the time sequence of pulses used for multimode spin-storage. (e) AFC efficiency measured as a function of AFC time $1/\Delta$, with bright input pulses and magnetic field of $\SI{1.35}{\milli\tesla}$ along $\text{\textbf{D}}_1$. The solid line indicates the exponential fit resulting in zero-time efficiency $\eta_0=(36\pm3)\%$ and effective coherence time $T_2^\text{AFC}=\SI{240\pm30}{\micro\second}$, see text for details. Error bars represent $95\%$ confidence intervals. More information on the experimental setup and pulse sequences can be found in the Methods section and in Supplementary Notes 1 and 3.}
	\label{fig:intro}
\end{figure*}

The realization of quantum repeaters~\cite{Briegel1998,Duan2001,Sangouard2011}, and more generally quantum networks, is a long-standing goal in quantum communication. It will enable long-range quantum entanglement distribution, long-distance quantum key distribution (QKD), distributed quantum computation and quantum simulation~\cite{Kimble2008}. Many schemes of quantum repeaters rely on the heralding of entanglement between quantum nodes in elementary links~\cite{Cabrillo1999a,Duan2001}, followed by local swapping gates~\cite{Briegel1998} to extend the entanglement. The introduction of atomic ensembles as repeater nodes, and the use of linear optics for the entanglement swapping, stems from the seminal DLCZ proposal~\cite{Duan2001}. A key advantage of atomic ensembles is their ability to store qubits in many modes through multiplexing~\cite{Nunn2008,Afzelius2009a,Sinclair2014,Laplane2015,Parniak2017,Yang2018,Heller2020}, which is crucial for distributing entanglement efficiently and with practical rates~\cite{Simon2007}.

Rare-earth-ion (RE) doped crystals provide a solid-state approach for ensemble-based quantum nodes. RE doped crystals can provide multiplexing in different degrees of freedom~\cite{Usmani2010,Sinclair2014,Laplane2015,Seri2017,Yang2018,Seri2019}, efficient storage~\cite{Sabooni2013,Hedges2010}, long optical coherence times~\cite{Equall1994,Equall1995}  and long coherence times of hyperfine states~\cite{Heinze2013,Zhong2015,Businger2020, Holzaepfel2020} that allows long-duration and on-demand storage of optical quantum states.  Long optical coherence times, in combination with the inhomogeneous broadening, offer the ability to store many temporal modes~\cite{Simon2007,Afzelius2009a}. Repeater schemes based on both time and spectral multiplexing schemes have been proposed~\cite{Simon2007,Sinclair2014}. Here we focus on repeaters employing time-multiplexing and on-demand read-out in time~\cite{Simon2007}, which require the long storage times provided by hyperfine states~\cite{Wu2020}.

The longest reported spin storage time of optical states with mean photon number of around 1 in RE doped solids is about 1 ms in \euiso{}~\cite{Jobez2015}. However, even near-term quantum repeaters spanning distances of 100 km or above would certainly require storage times of at least 10 ms, and more likely of hundreds of ms~\cite{Wu2020}. A particular challenge of long duration quantum storage in RE systems is noise introduced by the application of the dynamical decoupling (DD) sequences that are required to overcome the inhomogeneous spin dephasing~\cite{Jobez2015} and the spectral diffusion~\cite{Zhong2015, Holzaepfel2020}. To reduce the noise one can apply error-compensating DD sequences~\cite{Cruzeiro2016}, or increase the spin coherence time by applying magnetic fields to reduce the required number of pulses~\cite{Fraval2004,Zhong2015,Etesse2021}.

In this article we report on an atomic frequency comb (AFC) spin-wave memory in \euiso{}, in which we demonstrate storage of 6 temporal modes with mean photon occupation number $\mu_\text{in} = 0.711\pm0.006$ per mode for a duration of \SI{20}{\milli\second} using a XY-4 DD sequence with 4 pulses. The output signal-to-noise (SNR) ratio is $7.4\pm0.5$, for an internal storage efficiency of $\eta_\text{s} = 7\%$, which excludes the contribution of losses due to the optical path between the memory output and the detector. These results represent a 40-fold increase in qubit storage time with respect to the longest photonic qubit storage in a solid-state device~\cite{Jobez2015}. The improvement in storage time is due to the application of a small magnetic field of \SI{1.35}{\milli\tesla}, see also~\cite{Holzaepfel2020,Etesse2021}, which increases the spin coherence time with more than an order of magnitue while simultaneously resulting in a Markovian spin diffusion that can be further suppressed by DD sequences. By applying a longer DD sequence of 16 pulses (XY-16) we demonstrate storage with $\mu_\text{in} = 1.062\pm0.007$ per mode for a duration of \SI{100}{\milli\second}, with a SNR of $2.5\pm0.2$ and an efficiency of $\eta_\text{s}$ = ($2.60\pm0.02$)\%. In addition we stored two time-bin qubits for \SI{20}{\milli\second} and performed a quantum state tomography of the output state, showing a fidelity of $F=(85\pm2)\%$ for $\mu_\text{in}$ = $0.92\pm0.04$ photons per qubit. To analyse the qubit we propose a composite adiabatic control pulse that projects the output qubit on superposition states of the time-bin modes.
The current limit in storage time is technical, due to heating effects in the cryo cooler caused by the high power of the DD pulses and the duty cycle of the sequence.
The measured spin coherence time as a function of the DD pulse number $n_\text{p}$ follows closely the expected $n_\text{p}^{2/3}$ dependence, which suggests that considerably longer storage times are within reach with some engineering efforts.

\section*{Results}
\subsection*{The \euiso{} system}
The platform for our quantum memory is a \euiso{} crystal with an energy structure at zero magnetic field as in Figure~\ref{fig:intro}a. The excited and ground states can be connected via optical transitions at about \SI{580}{\nano\meter}~\cite{Koenz2003}. The quadrupolar interaction due to the effective nuclear spin $I=5/2$ of the \eu{} ions generates three doublets in both the ground and excited states, separated by tens of MHz. This structure allows to choose a $\Lambda$-system with a first ground state $\ket{\text{g}}$ into which the population is initialized, connected to an excited state $\ket{\text{e}}$ for optical absorption of the input light, and a second ground state $\ket{\text{s}}$ for on-demand long-time storage.

The full AFC-spin wave protocol~\cite{Afzelius2009a,Afzelius2010}, sketched in Figure~\ref{fig:intro}d, begins by initializing the memory so to have a comb-like structure in the frequency domain with periodicity $\Delta$ on $\ket{\text{g}}$ and an empty $\ket{\text{s}}$ state, via an optical preparation beam (see Figure~\ref{fig:intro}b). The initialization step closely follows the procedure outlined by Jobez et al.~\cite{Jobez2016}. The photons to be stored are sent along the input path and are absorbed by the AFC on the $\ket{\text{g}}\leftrightarrow\ket{\text{e}}$ transition, leading to a coherent superposition in the atomic ensemble. The AFC results in a rephasing of the atoms after a duration $1/\Delta$, while normally they would dephase quickly due to the inhomogeneous broadening. Before the AFC echo emission, the excitation is transferred to the storage state $\ket{\text{s}}$ via a strong transfer pulse. The radio-frequency (RF) field at \SI{46.18}{\mega\hertz} then dynamically decouples the spin coherence from external perturbations and compensates for the spin dephasing induced by the inhomogeneous broadening of the spin transition $\ket{\text{g}}\leftrightarrow\ket{\text{s}}$. In our particular crystal, the shape of the spin transition absorption line is estimated to be Gaussian with a width of about \SI{60}{\kilo\hertz} (Supplementary Note 2). A second strong optical pulse transfers the coherent atoms back into the $\ket{\text{e}}$ state, after which the AFC phase evolution concludes with an output emission along $\ket{\text{e}}\rightarrow\ket{\text{g}}$. 

To implement the memory scheme, a coherent and powerful laser (\SI{1.8}{\watt}) at \SI{580}{\nano\meter} is generated by amplifying and frequency doubling a \SI{1160}{\nano\meter} laser that is locked on a high-finesse optical cavity~\cite{Jobez2014}. The \SI{580}{\nano\meter} beam traverses a cascade of bulk acousto-optic modulators (AOM), each controlling an optical channel of the experiment, namely optical transfer, memory preparation, filter preparation and input. The optical beams and the main elements of the setup are represented in Figure~\ref{fig:intro}b, c. A memory and two filtering crystals are cooled down in the same closed-cycle helium cryostat to $\sim\SI{4}{\kelvin}$, placed on two levels of a single custom mount. The $\SI{1.2}{\centi\meter}$ long memory crystal is enveloped by a coil of the same length to generate the RF field. Another larger coil is placed outside the cold chamber and used to generate a static magnetic field.
After the cryostat, the light in the input path can be detected either by a linear Si photodiode for experiments with bright pulses, or by a Si single photon avalanche diode (SPAD) detector for weak pulses at the single photon-level. For photon counting it is necessary to use a filtering setup (Figure~\ref{fig:intro}c) to block any scattered light and noise generated by the second transfer pulse. Another AOM acts as a temporal gate, before passing the beam through two filtering crystals that are optically pumped so to have a transmission window around the input photon frequency and maximum absorption corresponding to the transfer pulse transition.

The \euiso{} crystals are exposed to a small static magnetic field along the crystal $\text{\textbf{D}}_1$ axis~\cite{Li1992}. At zero magnetic field, the protocol enabled to achieve storage of multiple coherent single photon-level pulses up to about \SI{1}{\milli\second}~\cite{Jobez2015}. However, it has been shown that even a weak magnetic field can increase the coherence lifetime~\cite{Equall1994,Alexander2007b,Arcangeli2014,Holzaepfel2020}, which motivated us to use a $\sim\SI{1.35}{\milli\tesla}$ field along the $\text{\textbf{D}}_1$ axis~\cite{Etesse2021}.

\subsection*{Spin-wave AFC}

The AFC spin-wave memory consists of three distinct processes: the AFC echo, the transfer pulses and the RF sequence, and each process introduces a set of parameters that will need to be optimized globally in order to achieve the best possible SNR, multimode capacity and storage time. Below we briefly describe some of the constraints leading to the particular choice of parameters used in these experiments.

The maximum AFC spin-wave efficiency is limited by the AFC echo efficiency for a certain $1/\Delta$, which typically decreases exponentially as a function of $1/\Delta$. We can define an effective AFC coherence lifetime $T_{2}^{\text{AFC}}$ and efficiency $\eta_{\text{AFC}}$ as $\eta_{\text{AFC}}=\eta_0\,\exp(-4/(\Delta T_2^{\text{AFC}}))$~\cite{Jobez2016}, where $\eta_0$ depends on the optical depth and the AFC parameters. With an external magnetic field of $\SI{1.35}{\milli\tesla}\parallel~D_1$, we obtained $T_2^{\text{AFC}}=\SI{240\pm30}{\micro\second}$ with an extrapolated zero-time efficiency of $\eta_0=\SI{36\pm3}{\%}$, see Figure~\ref{fig:intro}e. The $\eta_0$ efficiency is consistent with the initial optical depth of $6$ in our double-pass input configuration (each pass provides an optical depth of about $3$). The effect of the field-induced Zeeman split on the AFC preparation process is discussed in detail Ref~\cite{Etesse2021}. In short no adverse effects of the comb quality is expected when the comb periodicity is a multiple of the excited state splitting, provided more than two ground states are available for optical pumping as in \eu{}:\yso{}, while other periodicities can lead to a lower AFC efficiency. In Figure~\ref{fig:intro}e the modulation period of about $\SI{25}{\micro\second}$ indeed corresponds to the excited state splitting of $\SI{41.4}{\kilo\hertz}$. The exponential decay implies that there is a trade-off between the memory efficiency (favoring short $1/\Delta$) and temporal multi-mode capacity (favoring long $1/\Delta$). In addition we must consider the shortest input duration that can be stored, which is limited by the effective memory bandwidth.

The optical transfer pulses should ideally perform a perfect coherent population inversion between states $\ket{\text{e}}$ and $\ket{\text{s}}$, uniformly over the entire bandwidth of the input pulse. Efficient inversion with a uniform transfer probability in frequency space can be achieved by adiabatic, chirped pulses~\cite{Minar2010}. Here we employ two HSH pulses proposed by Tian et al.~\cite{Tian2011a}, which are particularly efficient given a limitation in pulse duration. For a fixed Rabi frequency the bandwidth of the pulse can be increased by increasing the pulse duration~\cite{Minar2010}, which however reduces the multimode capacity of the AFC spin-wave memory. 

Considering as a priority to preserve the storage efficiency while still being able to store several time modes, we set $1/\Delta=\SI{25}{\micro\second}$, corresponding to the first maximum (with efficiency $28\%$) on the AFC echo decay curve in Figure~\ref{fig:intro}e. Given the $1/\Delta$ delay, we optimized the HSH control pulse duration, leading to a bandwidth of \SI{1.5}{\mega\hertz} for a HSH pulse duration of \SI{15}{\micro\second}. The remaining \SI{10}{\micro\second} were used to encode $6$ temporal modes, giving a mode duration of $T_\text{m} = \SI{1.65}{\micro\second}$. Each mode contained a Gaussian pulse with a full-width at half-maximum of about 700 ns.

The RF sequence compensates for the inhomogeneous spin dephasing and should ideally reduce the spectral diffusion due to spin-spin interactions through dynamical decoupling (DD)~\cite{Viola1998,Lange2010,Medford2012,Holzaepfel2020}. However, effective dynamical decoupling requires many pulses, with pulse separations less than the characteristic time of the spin fluctuations. Pulse errors can then introduce noise at the memory output~\cite{Jobez2015}, which in principle can be reduced by using error-compensating DD sequences~\cite{Cruzeiro2016}. In practice, however, other factors such as heating of the crystal due to the intense RF pulses limit the effectiveness of such sequences, and noise induced by the RF sequence is the main limitation in SNR of long-duration AFC spin-wave memories~\cite{Jobez2015,Laplane2015,Laplane2017}.

\subsection*{Characterization with bright pulses}

\begin{figure}
	\includegraphics[width=\columnwidth]{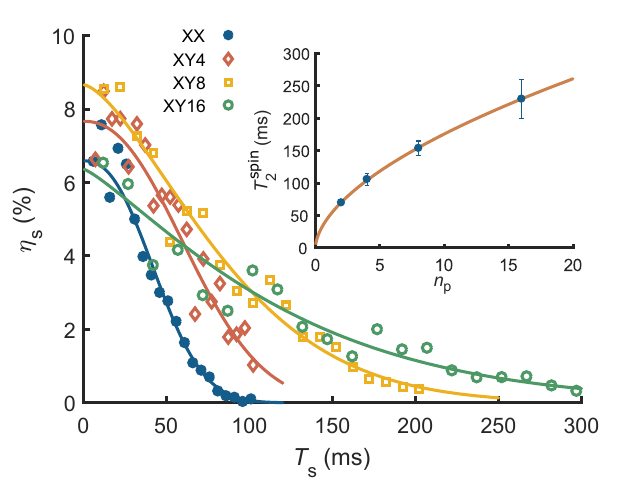}
	\caption{\textbf{Spin storage coherence time} Spin storage efficiency as a function of storage time for four different dynamical decoupling sequences. The solid lines are fits of the Mims model with $\eta_\text{s}(0)$, $T_2^\text{spin}$ and $m$ as free parameters (see text for details). Inset: spin effective coherence time as a function of the number of pulses $n_\text{p}$ in the DD sequence. The solid line is a fit to a power law as described in the main text. Error bars indicate a \SI{95}{\percent} confidence interval. See Supplementary Table 1 for details on the $T_2^\text{spin}$ data.}
	\label{fig:spin_curves}
\end{figure}

We first present a characterization of the memory using bright input pulses and a linear Si photodiode, implementing four decoupling sequences with a number of pulses ranging from a minimum of 2 to a maximum of 16. Figure~\ref{fig:spin_curves} displays the resulting efficiency decay curves as a function of the time $T_\text{s}$ spent by the atoms in the spin transition, which corresponds to the time difference between the two optical transfer pulses. The solid lines show fits obtained from a Mims model, which takes into account the effect of spectral diffusion~\cite{Mims1968} according to the relation $\eta_\text{s}(T_\text{s})=\eta_\text{s}(0) \,\exp[-2(\, T_\text{s}/T_2^\text{\,spin})^m]$, where $T_2^\text{\,spin}$ is the effective spin coherence time, and $m$ the Mims factor. More details can be found in Supplementary Note 3.

\begin{figure*}[htp]
	\includegraphics[width=\linewidth]{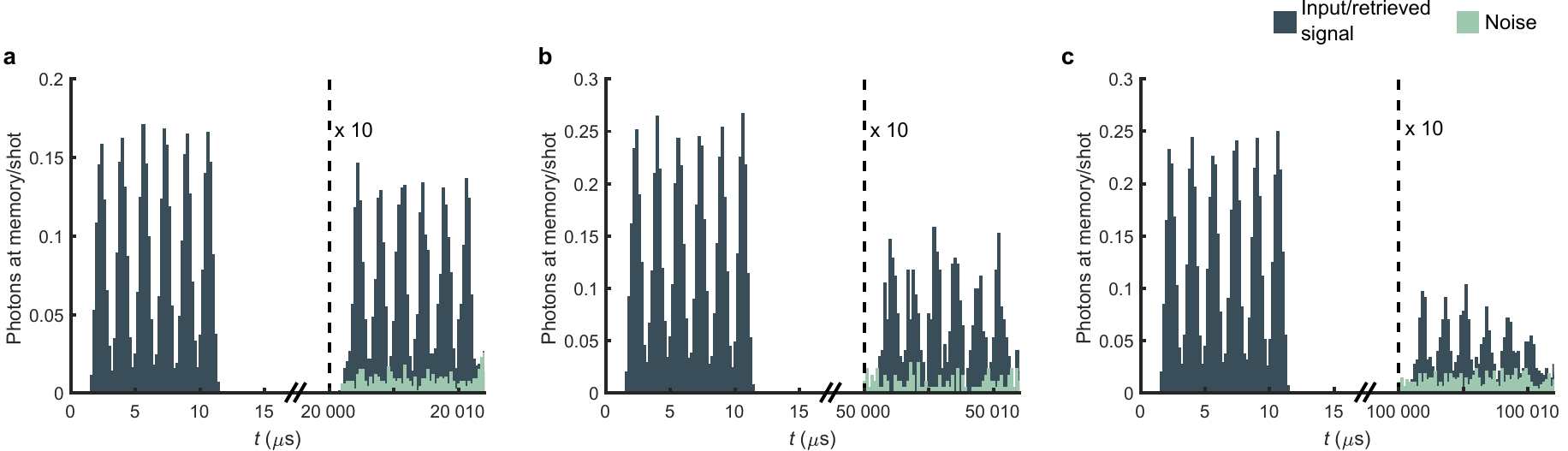}
	\caption{\textbf{Single photon-level spin storage} Examples of spin storage at \SI{20}{\milli\second} (a), \SI{50}{\milli\second} (b) and \SI{100}{\milli\second} (c). The dark blue histogram shows the input pulses (left of each figure) and the retrieved signal. Light green histograms display the noise floor. Each signal peak is a the center of a \SI{1.65}{\micro\second} time mode. The measured mean photon number in each input mode was close to 1, within the statistical variations. The mean photon number averaged over all 6 modes is given in Table~\ref{tab:storage_results}. See Methods and Supplementary Tables 2, 3 and 5 for more details.}
	\label{fig:singlephoton_storage}
\end{figure*}
We extracted effective coherence times of $70\pm2$, $106\pm9$, $154\pm11$ and $\SI{230\pm30}{\milli\second}$ respectively for XX, XY-4, XY-8 and XY-16 sequences, which show a clear decoupling effect as more pulses are added. This is also confirmed by the expected change of $T_2^\text{\,spin}$ as a function of $n_\text{p}$ visible in the inset of Figure~\ref{fig:spin_curves}, which closely follows a power-law relation $T_2^\text{\,spin}(n_\text{p})=T_2^\text{\,spin}(1)\,n_\text{p}^{\gamma_\text{p}}$ with $\gamma_\text{p}=0.57\pm0.03$ and $T_2^\text{\,spin}(1)=\SI{47\pm2}{\milli\second}$, as expected for a Ornstein-Uhlenbeck spectral diffusion process~\cite{Klauder1962,Lange2010,Medford2012}. A similar scaling was obtained in Holz\"apfel et al.~\cite{Holzaepfel2020}, using a slightly different experimental setup, magnetic field and $\Lambda$-system, which indicates that much longer storage times could be achieved. However, adding more pulses for the same storage times introduces additional heating, causing temperature-dependent frequency shifts of the optical transition~\cite{Koenz2003,Thorpe2013}. This technical issue could be addressed in the future by optimizing the heat dissipation in proximity of the crystal. The extrapolated zero-time efficiencies vary between $6$ and $9\%$, and the data appears relatively scattered around the fitted curves for the longer decoupling sequences. These two observations might be a sign of the presence of beats originating in the different phase paths available to the atoms during storage, due to the small Zeeman splitting of the ground state doublets in this regime of weak magnetic field. Similar effects have been shown in a more detailed model of interaction between a system with splittings smaller than the RF pulses chirp~\cite{Etesse2021}.

\subsection*{Single-photon level performance}

We now discuss the memory performance at the single photon level. The dark histograms in Figure~\ref{fig:singlephoton_storage} show three examples of spin storage outputs with their respective input modes for reference. The lighter histograms show the noise background, measured while executing the complete memory scheme without any input light (see Methods section for details). This noise floor, when integrated over the mode size $T_\text{m}$, gives us the noise probability $p_\text{N}$. When compared to the sum of the counts in the retrieved signal in the mode, the summed noise count is well below the retrieved signal for all the storage times here reported. We used an XY-4 type of RF sequence for storage at \SI{20}{\milli\second}, XY-8 for \SI{50}{\milli\second} and XY-16 for \SI{100}{\milli\second}. Table~\ref{tab:storage_results} summarizes the relevant results, in particular with SNR values ranging from $7.4$ to $2.5$ for \SI{20}{\milli\second} and \SI{100}{\milli\second} respectively. The average input photon number per time mode $\mu_\text{in}$ is close to $1$ in all cases, although it varies slightly. To account for this, an independent figure of merit is the parameter $\mu_1=p_\text{N}/\eta$, which corresponds to the average input photon number that would give an SNR of $1$ in output~\cite{Gundogan2015}. Since it scales as the inverse of the efficiency~\cite{Jobez2015}, it increases with storage time, but for all cases studied here it is well below $1$. 

The noise probability $p_\text{N}$ varied from $7 \cdot 10^{-3}$ to $11 \cdot 10^{-3}$, see Table~\ref{tab:storage_results}, similar to previous experiments~\cite{Jobez2015}. An independent noise measurement at 20 ms showed that the XX, XY-4 and XY-8 resulted in almost identical noise values of $p_\text{N} = 7.4\cdot 10^{-3}$, $8.1\cdot 10^{-3}$ and $8.6\cdot 10^{-3}$ (error $\pm0.3\cdot 10^{-3}$), respectively. This shows that pulse area errors are effectively suppressed by the higher order DD sequences, and that the read-out noise is caused by other types of errors, which at this point are not well understood.

\begin{table*}[htp]
	\centering
	\begin{tabular}{c@{\qquad}c@{\qquad}c@{\qquad}c@{\qquad}c@{\qquad}c}
		{$T_\text{s}$}	& {$\mu_\text{in}$}	&$p_\text{N}$				& {$\eta$}			& {SNR} 		&{$\mu_1$}\\
		(ms)	&					&					& (\%)				&				&\\
		\midrule[0.5pt]
		$20$	& $0.711\pm0.006$	&$0.0073\pm0.0012$	& $7.39\pm0.04$		& $7.4\pm0.5$ 	& $0.098\pm0.002$\\[2pt]
		$50$	& $1.21\pm0.01$		&$0.009\pm0.002$	& $4.37\pm0.04$		& $5.6\pm0.7$ 	& $0.218\pm0.008$\\[2pt]
		$100$	& $1.062\pm0.007$ 	&$0.0110\pm0.0015$	& $2.60\pm0.02$		& $2.5\pm0.2$ 	& $0.445\pm0.008$\\
	\end{tabular}
	\caption{\textbf{Summary of single photon-level storage.} Values of input mean photon number $\mu_\text{in}$, storage efficiency $\eta$, signal-to-noise ratio SNR and the equivalent mean input photon number for SNR=$1$, $\mu_1$, for different spin-wave storage times $T_\text{s}$. The reported values are averages over the 6 temporal modes, and the value for each mode stems from the summed counts over the mode size $T_\text{m}$. See Methods and Supplementary Tables 2, 3 and 5 for more details.}
	\label{tab:storage_results}
\end{table*}

If compared with the efficiencies measured with bright pulses in Figure~\ref{fig:spin_curves}, the storage efficiency measured at the single photon-level is noticeably lower for $50$ and \SI{100}{\milli\second}. We believe this is due to the long measurement times required for accumulating the necessary statistics, which exposes the experiment to long-term fluctuations affecting optical alignment in general and specifically fiber coupling efficiencies. Nonetheless, our results show that our memory is capable of storing successfully multiple time modes at the single photon level, with a SNR that is in principle compatible with storage of quantum states~\cite{Jobez2015} for up to \SI{100}{\milli\second}. More information on the memory parameter estimations from the data can be found in Supplementary Table 2.

\subsection*{Time-bin qubit storage}

\begin{figure*}[htp]
	\includegraphics[width=\linewidth]{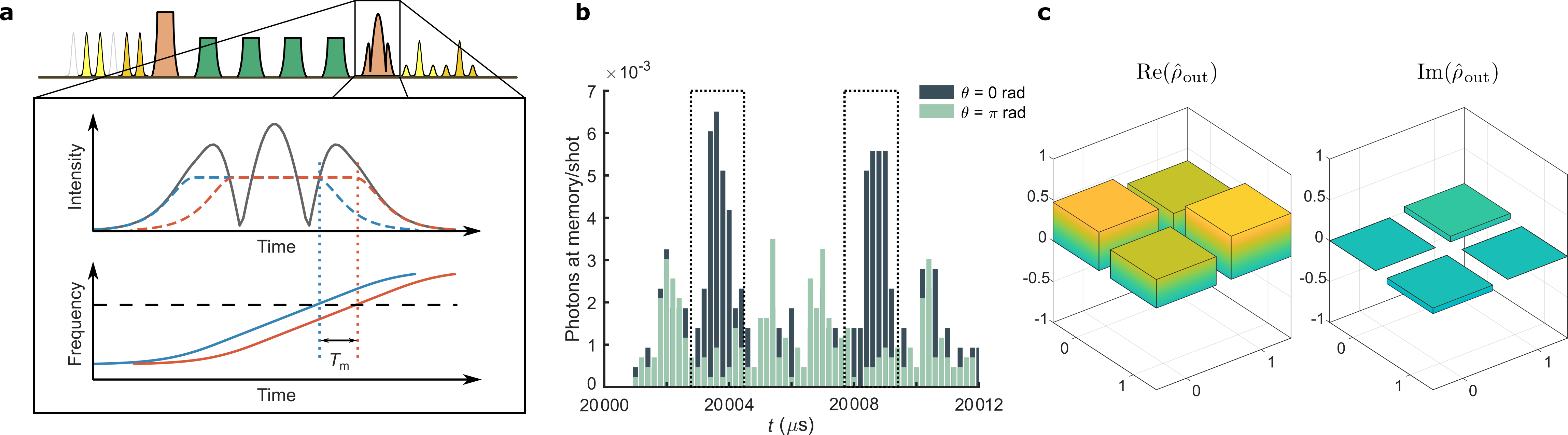}
	\caption{\textbf{Qubit tomography} (a) Projections for the tomography are implemented by choosing the appropriate pulse profile for the second transfer pulse in the storage sequence. For projections in $\sigma_\mathrm{x}$ and $\sigma_\mathrm{y}$ the second transfer pulse takes the form of a composite HSH, obtained from the sum of the fields of two chirped pulses that are temporally shifted by the width of one time bin ($T_\text{m}$) relatively to each other. In the top figure, the envelopes of the two partial pulses are shown with dashed lines and the resulting composite pulse with a solid line. As a consequence, each specific frequency of the AFC (e.g. the dashed line in the bottom figure) is addressed at two different times separated by $T_\text{m}$, as indicated by the crossings with the two solid lines (b) Two examples of projection of the output state onto $\ket{+}$ (dark histogram) and $\ket{-}$ (light histogram) eigenstates of the $\sigma_\mathrm{x}$ operator. The overlap is proportional to the amplitude in the interference bins (dotted boxes). The input state $\psi_\text{in}$ was prepared in the $\ket{+}$ eigenstate of $\sigma_\mathrm{x}$ (c) Reconstructed density matrix $\hat{\rho}_\text{out}$ of the output state. See Supplementary Note 4 for more details.}
	\label{fig:qubit}
\end{figure*}

To characterize the quantum fidelity of the memory, we analyzed the storage of time-bin-encoded qubits. Both qubits were prepared in the ideally pure superposition state $\psi_\text{in} = 1/\sqrt{2}\,(\ket{\text{E}} + \ket{\text{L}})$, where $\ket{\text{E}}$ and $\ket{\text{L}}$ represent the early and late time modes of each qubit. Exploiting our 6-modes capacity, we encoded the components $\ket{\text{E}}$ and $\ket{\text{L}}$ of the first qubit into the temporal modes 2 and 3 respectively, and similarly for the second qubit in modes 5 and 6.

To perform a full quantum tomography of the memory output state, represented by the density matrix $\rho_\text{out}$, one needs to be able to perform measurements of the observables represented by the Pauli matrices $\sigma_\mathrm{x}$, $\sigma_\mathrm{y}$ and $\sigma_\mathrm{z}$. The observable $\sigma_\mathrm{z}$ can simply be measured using histogram traces as shown in Figure~\ref{fig:singlephoton_storage}. The $\sigma_\mathrm{x}$ and $\sigma_\mathrm{y}$ observables can be measured by making two partial read-outs of the memory \cite{Staudt2007a,Gundogan2013,Gundogan2015}, separated by the qubit mode spacing $T_\text{m}$, where each partial transfer pulse should ideally perform a $50\%$ transfer as both modes are emitted after the second transfer pulse. In the past, this has been achieved by using two distinct, shorter transfer pulses~\cite{Gundogan2013,Gundogan2015,Ma2021}, separated by $T_\text{m}$, which in practice can reduce the efficiency below the ideal 50\% transfer~\cite{Gundogan2015}. This is particularly true for long adiabatic, chirped pulses, which would then need to be severely shortened to produce two distinct pulses separated by the mode spacing $T_\text{m}$. In addition the first control pulse would need to be reduced in duration as well, as the chirp rate of all the transfer pulses should be the same~\cite{Minar2010}.

To overcome the efficiency limitation for qubit analysis based on partial read-outs with adiabatic pulses, we propose a composite pulse that can achieve the ideal 50\% partial transfer, independently of the pulse duration and mode separation. The composite HSH pulse (cHSH) is a linear sum of two identical adiabatic HSH pulses, with their centers separated in time by $T_\text{m}$. The cHSH has a characteristic amplitude oscillation due to the interference of the two chirps, see Figure~\ref{fig:qubit}a. Intuitively, one can think of each specific frequency within the AFC bandwidth as being addressed twice by the cHSH, once by each component, at two distinct times separated exactly by $T_\text{m}$, despite the fact that the whole cHSH pulse itself is much longer than $T_\text{m}$. As a consequence, the addressed atomic population partially rephases after the pulse at two times separated by $T_\text{m}$. An alternative method for analysing qubits consists in using an AFC-based analyser in the filtering crystal~\cite{Jobez2015a,Kutluer2019}. However, we observed that the SNR was deteriorated when using the same crystal as both filtering and analysing device. The cHSH-based analyser resulted in a significantly better SNR after the filters, yielding a higher storage fidelity.

The phase difference $\theta$ between the two cHSH components sets the measurement basis, where $\theta=0$ ($\theta=\mathrm{\pi}/2$) and $\theta=\mathrm{\pi}$ ($\theta=3\mathrm{\pi}/2$) projects respectively on the $\ket{+}$ and $\ket{-}$ eigenstates of $\sigma_\mathrm{x}$ ($\sigma_\mathrm{y}$), encoded in the early-late time modes basis as $1/\sqrt{2}\,(\ket{\text{E}}+ \mathrm{e}^{\mathrm{i}\theta}\ket{\text{L}})$. Note that this type of analyser can only project onto one eigenstate of each basis, hence two measurements are required per Pauli operator. Figure \ref{fig:qubit}b shows the histograms corresponding to the two $\sigma_\mathrm{x}$ projections. 

After measuring the expectation value of all three Pauli operators, we can reconstruct the full quantum state $\hat{\rho}_\text{out}$ using direct inversion~\cite{Schmied2016}, after verifying that the corresponding state matrix is indeed physical. We hence derive a fidelity of $F=\SI{85\pm2}{\percent}$, averaged over the two qubits. Raw counts and the resulting expactation values can be found in Supplementary Table 4, with corresponding numbers of experiment repetitions in Supplementary Table 6. The average number of photons per qubit was $\mu\text{in}$ = 0.92$\pm$0.04 and the reconstructed density matrix $\hat{\rho}_\text{out}$ is shown in Figure~\ref{fig:qubit}c. The purity of the reconstructed state is $P = \SI{76\pm 3}{\percent}$, which limits the maximum achievable fidelity in absence of any unitary errors to $87\%$. This indicates that the fidelity is limited by white noise generated by the RF sequence at the memory read-out. Another element supporting this conclusion is given by the fidelity measured with bright pulses, so that the noise is negligible (see Supplementary Note 4), yielding a value of \SI{96}{\percent}. We further note that the $\sigma_\mathrm{z}$ measurement yielded a SNR of $3.48\pm0.15$, which when scaled up to the single photon level in one time bin becomes about $7.0$. This is compatible with the value reported in Table~\ref{tab:storage_results} and would result in an upper bound on the fidelity of $F =(\text{SNR}+1)/(\text{SNR}+2) =(88.9 \pm 0.04)\%$ assuming a white noise model~\cite{Jobez2015}.

The measured qubit fidelity can be compared to different criteria for quantum storage. In this work we characterize the memory by storing qubits encoded onto weak coherent states. In this context Specht et al. \cite{Specht2011} introduced a classical fidelity limit by comparing to a measure-and-prepare strategy, where the memory inefficiency and multiphoton components of the states are exploited. Nevertheless, for the efficiency of $7.39\%$ and the mean qubit photon number of $\mu_\text{in} = 0.92$ the criterion gives a maximum classical fidelity of $81.2\%$ (see Supplementary Note 4), such that our qubit fidelity at \SI{20}{\milli\second} surpasses the classical limit. We can also consider future applications of the memory in terms of storing a qubit encoded onto a true single photon (Fock state), for which the classical limit is $F=2/3$~\cite{Massar1995}. For storage of true single photon qubits it can be shown that this limit can be surpassed provided that the probability $p$ of finding the photon before the memory is larger than the $\mu_{1}$ parameter (see Table~\ref{tab:storage_results}) \cite{Jobez2015,Laplane2015}. Recent quantum memory experiments in praesodymium-doped \yso{} has reached a heralding efficiency of $19\%$ of finding a true single photon before the memory~\cite{LagoRivera2021}, which with our $\mu_{1} = 0.098$ at \SI{20}{\milli\second} storage time would result in a theoretical qubit fidelity of about $75\%$. The fidelity can be improved by a combination of increasing current memory efficiency and single-photon heralding efficiency.

\section*{Discussion}

The results presented here demonstrate that long-duration quantum storage based on dynamical decoupling of spin-wave states in \euiso{} is a promising avenue. In terms of qubit storage, we observe a 40-fold increase in storage time with respect to the previous longest quantum storage of photonic qubits in a solid-state device~\cite{Laplane2015}. Currently, the storage time in \euiso{} is limited by the heating observed when adding more pulses in the decoupling sequence, which is a technical limitation, but the classical storage experiments by Holz\"apfel et al.~\cite{Holzaepfel2020} suggest that even longer storage times are within reach in \euiso{}. Solving the heating problem will also allow applying DD pulses in a rapid succession with fixed time separation, as done by Holz\"apfel et al.~\cite{Holzaepfel2020}, giving more flexifibilty in the read out time and reducing any deadtime of the memory. In general the implications of the timing of DD sequences have not yet been adressed in rate calculations of quantum repeaters. Our observation that DD sequences with more pulses did not generate more read-out noise is key to achieving longer storage times also at the quantum level. We also note that these techniques could be applied also to Pr$^{3+}$ doped \yso{} crystals, where currently quantum entanglement storage experiments are limited to about \SI{50}{\micro\second}~\cite{Rakonjac2021}. Another interesting avenue is to apply these techniques to extend the storage time of spin-photon correlations experiments~\cite{Laplane2017,Kutluer2019} in rare-earth-doped crystals. 

\section*{Methods}
\subsection*{Expanded setup}
The core of the setup consist of a closed-cycle pulsed helium cryostat with a sample chamber at a typical temperature of \SI{3.5}{\kelvin}. In the sample chamber, a custom copper mount holds the memory crystal, with dimensions 2.5~x~2.9~x~12.3 mm along the $(\text{\textbf{D}}_1,\text{\textbf{D}}_2,\text{\textbf{b}})$ axes~\cite{Li1992}, and a series of two filtering crystals with similar size. All these are \euiso{} crystals with a doping concentration of \SI{1000}{ppm}~\cite{Jobez2015}. Around the memory crystal, a copper coil generates the RF field to manipulate the atoms on their spin transitions. The coil is coupled to a resonator circuit, with resonance tuned on the \SI{46}{\mega\hertz} spin transition, which produces a Rabi frequency of \SI{120}{\kilo\hertz}, corresponding roughly to an AC field of amplitude \SI{12}{\milli\tesla}. Before the resonator, the RF signal is created by an arbitrary wave generator, and amplified with a \SI{100}{\watt} amplifier coupled to a circulator to redirect unwanted reflection from the resonator system.

The input beam is used to create the optical pulses to be stored, and it goes through the memory crystal twice with a waist diameter of \SI{50}{\micro\meter}. The memory preparation beam is overlapped with the input path with a larger spot size around \SI{700}{\micro\metre}, to ensure homogeneity of the preparation along the crystal length, with an incident angle of about \SI{1}{\degree}. The transfer beam is overlapped in a similar way, with a beam diameter at the waist of \SI{250}{\micro\meter}.

The AFC structure is prepared with a \SI{3}{\mega\hertz} total spectral width, which however is not fully exploited since the \SI{1.5}{\mega\hertz} bandwidth of the HSH transfer pulses limits the effective memory bandwidth. The width of the transparency and absorption windows in the filter crystals were of \SI{2}{\mega\hertz}, with a total optical depth of approximately $7.4$ in the absorption window through the two crystals. The relative spectral excinction ratio should thus be $\exp(7.4):1 = 1636:1$.

More information on the setup can be found in Supplementary Note 1 and Supplementary Note 2.

\subsection*{Photon counting and noise measurement}
The quantities $\mu_\text{in}$ and $p_\text{N}$ in Table~\ref{tab:storage_results}, correspond to average number of photons at the memory output for a single storage attempt. They are obtained by summing raw detections in modes of duration $T_\text{m}=\SI{1.65}{\micro\second}$, then dividing by the number of experiment repetitions, averaging over the 6 modes, and dividing by the detector efficiency $\eta_\text{D}=\SI{57}{\percent}$ and cryostat-to-detector path transmission (typically between \num{17} and \SI{20}{\percent}). The histograms in Figure~\ref{fig:singlephoton_storage} and Figure~\ref{fig:qubit}b are obtained in the same way for a binning resolution of \SI{200}{\nano\second}.

Noise photons at the memory read-out originate from excitation of ions from the $\ket{\text{g}}$ state to the $\ket{\text{s}}$ state during the DD sequence, due to imperfections of the RF pulses~\cite{Jobez2015,Cruzeiro2016}. These ions are then excited by the read-out transfer pulse and decay on the $\ket{\text{e}}$-$\ket{\text{g}}$ transition trough spontaneous emission. These noise photons are thus spectrally indistinguishable from the stored photons. The spontaneous character was verified by observing that its decay constant corresponds to the radiative lifetime $T_1$. Note that without RF manipulation the read-out noise is significantly reduced, i.e. the main SNR limitation in current spin-wave experiments is due to RF-induced photon noise.

The noise parameter $p_\text{N}$ indicates the probability of a noise photon being emitted by the memory during a time corresponding to the mode size $T_\text{m}=\SI{1.65}{\micro\second}$. For the spin storage data at $T_\text{s}=\SI{20}{\milli\second}$, visible in Figure~\ref{fig:singlephoton_storage}a and Table~\ref{tab:storage_results}, it is measured independently by blocking the input beam during the full storage sequence in the same time-modes in which the retrieved modes would be. A more detailed analysis per-mode and the exact number of repetitions for all experiments are reported respectively in Supplementary Table 3 and Supplementary Table 5.

To decrease the total acquisition time of the experiments at $T_\text{s}=50$ and $\SI{100}{\milli\second}$, we calculated the respective $p_\text{N}$ values reported in Table~\ref{tab:storage_results} from a $\sim\SI{225}{\micro\second}$ time window centered at about \SI{190}{\micro\second} after the first retrieved mode in the same dataset. By doing so, we exploited the fact that the noise floor is due to spontaneous emission with a decay time of \SI{1.9}{\milli\second}~\cite{Equall1994}, and can be considered uniform up to $\sim\SI{200}{\micro\second}$ after readout. This is confirmed experimentally on the \SI{20}{\milli\second} datasets, as the difference in $p_\text{N}$ calculated in the retrieved mode position of the data with input blocked correspond to the result of the procedure above within the Poissonian standard deviation. The histograms displaying noise in Figure~\ref{fig:singlephoton_storage}b and c are representative regions of the noise floor taken at about \SI{20}{\micro\second} after the retrieved modes in the same dataset.

All errors are estimated from Poissonian standard deviations on the raw detector counts and propagated considering the memory temporal modes as independent.

\section*{Data availability}
The data sets generated and/or analysed during the current study are available from the corresponding authors upon reasonable request.

\section*{Acknowledgements}

We acknowledge funding from the Swiss FNS NCCR programme Quantum Science Technology (QSIT), European Union Horizon 2020 research and innovation program within the Flagship on Quantum Technologies through GA 820445 (QIA) and under the Marie Sk\l{}odowska-Curie program through GA  675662 (QCALL).

We also thank Philippe Goldner and Alban Ferrier from Chimie ParisTech for fruitful discussions and for providing the crystals.

\section*{Author Contribution}
A.O., J.E. and M.A. conceived and planned the experiments, which were mainly carried out by A.O. and A.H.. A.O. set up most of the experiment, with contributions from J.E., and carried out the quantum memory characterization. A.H. implemented the qubit analysis method, with contributions from A.O. The manuscript was mainly written by A.O. and M.A., with contributions from all the authors. M.A. provided overall oversight of the project.

\section*{Competing interests}
The authors declare that there are no competing interests.


%


\end{document}